# Molecular transport through capillaries made with atomic scale precision

B. Radha[1*], A. Esfandiar[1], F. C. Wang[2], A. P. Rooney[3], K. Gopinadhan[1], A. Keerthi[1], A. Mishchenko[1], A. Janardanan[1], P. Blake[4], L. Fumagalli[1,4], M. Lozada-Hidalgo[1], S. Garaj[5], S. J. Haigh[3], I. V. Grigorieva[1], H. A. Wu[2], A. K. Geim[1*]

[1]School of Physics & Astronomy, University of Manchester, Manchester, M13 9PL, UK
[2]Chinese Academy of Sciences Key Laboratory of Mechanical Behavior & Design of Materials, Department of Modern Mechanics, University of Science & Technology of China, Hefei, Anhui 230027, China
[3]School of Materials, University of Manchester, Manchester, M13 9PL, UK
[4]National Graphene Institute, University of Manchester, Booth St. E, Manchester M13 9PL, UK
[5]Department of Physics, National University of Singapore, 117542 Singapore

**Nanoscale pores and capillaries have been studied intensively because of their importance in many natural phenomena and use in numerous applications[1]. Significant progress has been achieved in fabricating artificial capillaries with nanometer dimensions, which led to the emergence of new research areas including nanofluidics[2-4]. However, it remains extremely challenging to control capillary sizes at this spatial scale, especially because of surface roughness. Here we report ultimately narrow and smooth capillaries that can be viewed as if individual atomic planes were removed from a bulk crystal, leaving behind flat voids of a chosen height. The capillaries are fabricated by van der Waals assembly[5] of atomically flat materials using two-dimensional crystals[6] as spacers in between. To demonstrate the technology, we use graphene and its multilayers as archetypal two-dimensional materials and study water transport through channels ranging in height from a single atomic plane to many dozens of them. The unexpectedly fast flow (up to 1 m/s) is attributed to high capillary pressures (~1,000 bar) combined with large slip lengths. For channels that accommodate only a few layers of water, the flow exhibits a marked enhancement, which we associate with an increased structural order in nanoconfined water. Our work opens a venue for making capillaries and cavities with sizes tunable to angstrom precision and permeation properties controlled through a wide choice of atomically flat materials available for channel walls.**

There are two principal routes for making pores and capillaries with nanometer dimensions[7]. In the top-down approach, micro and nano fabrication techniques are employed, and channels down to 2 nm in average height were demonstrated[8]. However, progress along this route is fundamentally limited by surface roughness that is hard to reduce below a few nm using conventional materials and techniques[9]. In the alternative bottom-up approach, chemical synthesis is used. Despite its many advantages for scalable manufacturing, this approach provides limited flexibility, especially for making capillaries with dimensions larger than several Å. A notable exception is carbon and other nanotubes. Their advent has offered fascinating opportunities for studying mass transport through channels with nanometer diameters and atomically smooth walls[10-17]. However, despite a promise for new kinds of membrane and nanofluidic systems, it turned out to be very difficult to integrate nanotubes into macroscopic devices, which perhaps explains the continuing controversy about fast water transport through carbon nanotubes (CNTs). Conflicting findings from only a few experimental groups, who succeeded in studying their permeation properties[10-12,15,16], have been discussed intensively in theoretical literature[13,14,17] but with little further input from experiment. Furthermore, graphene has already attracted considerable attention as a basis material for making ultra-short nanopores[18-23]. Gas, liquid, ion and DNA transport through such pores have been reported. Unfortunately, the fundamental restrictions inherent to top-down and bottom-up techniques limit the ability to control exact diameters of graphene nanopores, too. Our approach described below exploits both atomic flatness of graphene and its atomic thinness. It allows micrometer-long channels with atomically smooth walls, somewhat similar to CNTs, but at the same time provides atomic-scale control of the channel's principal dimension (height) in combination with much of the flexibility offered by microfabrication techniques.

Fig. 1a explains the basic idea behind our nanocapillary devices. They consist of atomically-flat top and bottom graphite crystals that are separated by an array of spacers made from few-layer graphene. Such structures are fabricated by van der Waals (vdW) assembly using dry transfer techniques[5]. A free-standing Si nitride membrane with a rectangular hole serves as a mechanical support for the assembly. Figs. 1b-d show micrographs of some of our devices. For details of their fabrication, we refer to 'Making nanocapillary devices' in supporting information and Figs. S1 and S2. We denote our devices by the number $N$ of graphene layers used as spacers. The height $h$ of the cavity available for molecular transport can then be estimated as $N \cdot a$ where $a \approx 3.4$ Å is the interlayer distance in graphite, that is, the effective thickness of one graphene layer. All the capillaries reported here had the same channel width $w \approx 130$ nm, and 200 of them were incorporated within each device to increase molecular flow (Fig. 1). Their length $L$ varied from < 2 to ≈10 μm. Despite the large aspect ratios $w/h$, we found no sagging of the graphite walls, which would cause capillary closures (Fig. 1d and supporting information Fig. S3).

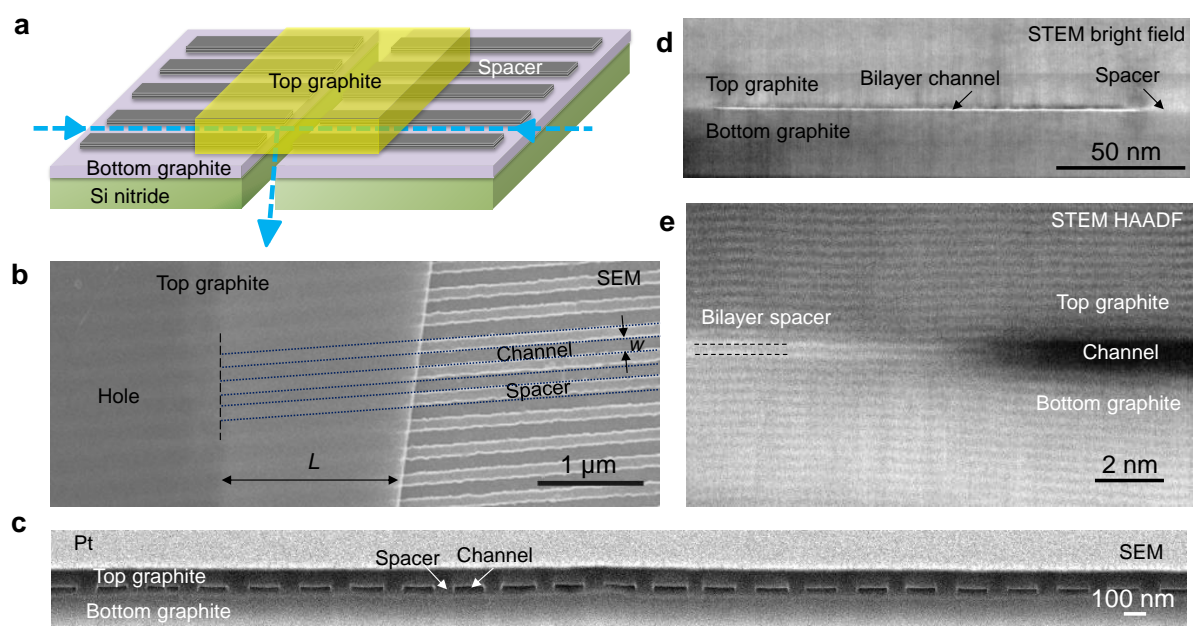

*Figure 1| Graphene capillary devices. a, Their principal schematic. The arrow indicates the flow direction used in all the experiments. b, Scanning electron microscopy (SEM) image of a trilayer device (top view). The spacers that are clearly seen in the area not covered by the top graphite can also be discerned underneath, running all the way to the hole etched in the bottom graphite. Three of the spacers are indicated by dotted lines and the edge of the hole by the dashed line. c, SEM micrograph of a cross-section of another device showing an array of capillaries with $h \approx 15$ nm. d, Cross-sectional bright field image of a bilayer capillary ($h \approx 7$ Å) in a scanning transmission electron microscope (STEM). e, High angle annular dark field (HAADF) image of the channel's edge. The lamellae for cross-sectional imaging were made by focused ion beam milling (see 'Visualization and characterization of graphene capillaries' in supporting information).*

Under ambient conditions, all surfaces are covered with various adsorbates including water and hydrocarbons[24], and it is not unreasonable to expect that nanocapillaries can be blocked by contamination introduced during fabrication or adsorbed from the air. Accordingly, we first checked whether our devices were open for gas and liquid transport. Figure S4 in supporting information shows that this was the case, and He permeated through the capillaries. We carried out He-leak tests for practically all the devices and found them normally open, except for monolayer capillaries ($N = 1$) which never exhibited any detectable permeation. Devices with larger $N$ gradually deteriorated and, after several days of measurements, often became blocked. We attribute this to a buildup of hydrocarbon contamination that creeps along surfaces and is present even under oil-free vacuum conditions in our He-leak experiments. On the other hand, if immersed in water, the capillaries showed much greater resilience. All the tested devices were found open (except for $N$

= 1, again) and exhibited ionic conductance scalable with their dimensions ('Ionic conductance' in supporting information and Fig. S5). If kept in water, the capillaries did not get blocked for months and could be repeatedly measured. The fact that such artificial channels with the height down to angstrom scale allow studies of molecular transport under normal experimental conditions is perhaps the most important finding of this work.

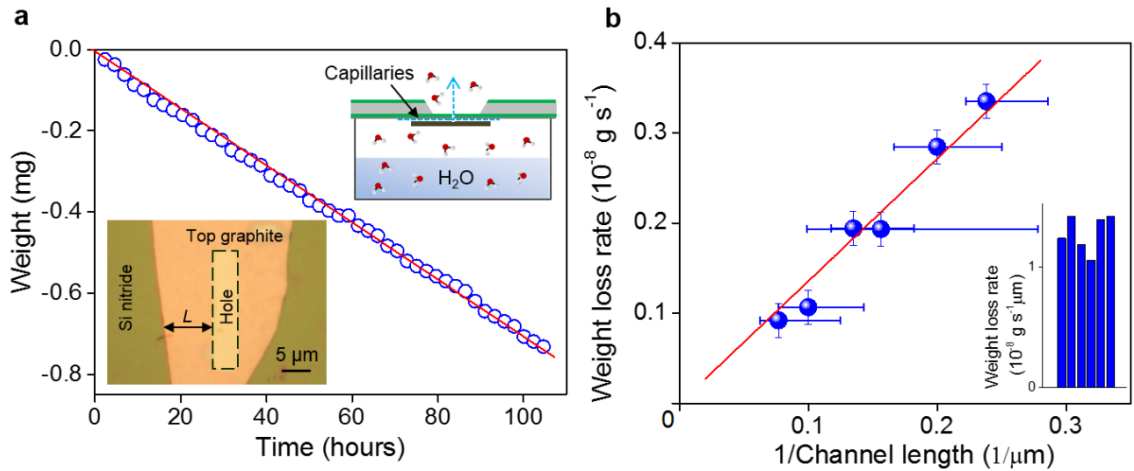

*Figure 2| Water permeation through graphene nanocapillaries. **a**, Weight loss due to water evaporation through one of our trilayer devices. Bottom inset: Optical image of this particular device that has 200 parallel channels with L ranging from 3.6 to 10.1 μm. The top inset illustrates a basic schematic of our gravimetric measurements. **b**, Weight loss rate, Q, measured for 6 trilayer devices with different effective $\tilde{L}$ (symbols). The error bars indicate the range of L within each device. Inset: Same data but normalized by $\tilde{L}$. The bars' heights correspond to the measured Q.*

Because of the intense interest in nanoconfined water and taking into account the high stability of our devices in water, we focus below on their properties with respect to water transport. To this end, we used precision gravimetry as shown schematically in Fig. 2a and described in detail in supporting information. Briefly, we measured a weight loss from a miniature container filled with water and sealed with a Si nitride chip incorporating a nanocapillary device (Fig. 2a, inset and supporting information Fig. S6). An example of such measurements is shown in Fig. 2a. The slope of the measured curve yields the water evaporation rate, $Q$. Because the total cross section of our devices is typically < 0.1 μm$^2$, measurements with μg precision over several days are required to achieve accurate readings of $Q$. Fig. 2b shows $Q$ observed for six devices with the same height ($N$ = 3) but different $L$. Within our accuracy, $Q$ was found to vary proportionally to $1/\tilde{L}$ = <1/L>, where $\tilde{L}$ is the effective average length with respect to a viscous flow, and < > denotes averaging over contributions from channels with different $L$ [see eq. (1) below]. This dependence on $L$ indicates that the observed evaporation rate is limited by water flow through capillaries, in agreement with other experimental observations described in supporting information 'Gravimetric measurements'. Reproducibility of our gravimetry results can be judged from the scatter in the inset of Fig. 2b where $Q$ for the trilayer devices are normalized by their $\tilde{L}$. All the trilayer devices show practically the same $Q \approx 10^{-8}$ g s$^{-1}$ normalized for 1 μm effective length, which translates into a flow velocity of $\approx 0.1$ m s$^{-1}$ for the shortest device in Fig. 2b ($\tilde{L} \approx 4$ μm). As a control, we fabricated devices following exactly the same fabrication procedures but without graphene spacers ($N$ = 0), in which case no weight loss could be detected. In addition, we tested our gravimetric setup using micrometer apertures made in Si nitride membranes and found evaporation rates that agree well with those expected from theory (supporting information 'Gravimetric measurements').

Having proven the accuracy and reproducibility of our measurements using trilayer devices, we investigated how the capillary flow depended on $N$ using more than 30 different devices. Fig. 3 shows that, as $h$ decreases from $\approx 10$ nm (maximum height in our gravimetric experiments), $Q$ also decreases, as generally expected. However, for $h < 2$ nm, $Q$ unexpectedly shoots up by more than an order of magnitude with respect to the trend exhibited by large-$N$ capillaries, and a profound peak appears at $N$ = 4–5 (Fig. 3b). Devices with monolayer spacers exhibited no detectable weight loss, similar to the case of $N$ = 0 and in agreement with our He-leak and ion-conductance tests.

To understand the observed behavior, it is important to note that, if the container was weighed upside down so that the liquid was in direct contact with capillaries' entries, exactly the same $Q$ were recorded as in the upright position (supporting information 'Gravimetric measurements'). This is not surprising because, at 100% humidity inside the container and the contact angle of water on graphite[25] $\phi \approx$ 55–85°, our graphene channels should be filled with the liquid due to capillary condensation[1,25]. Furthermore, the observed permeation rates for water are at least three orders of magnitude greater than those for He gas driven by a pressure of 23 mbar (difference in water vapor pressures inside and outside the container), which rules out the possibility that water permeates through our capillaries as a vapor (supporting information 'Helium-leak testing'). In additional experiments, we pressurized our containers at ≈1.5 bar (close to the maximum pressure that our membranes could withstand) but no difference in $Q$ could be discerned (supporting information 'Gravimetric measurements'). This shows that the observed evaporation is driven not by the small difference in vapor pressures but by a much higher pressure. We assign the latter to a capillary pressure $P$ that can be approximated[1,26] as $P_0 + \Pi = 2\sigma\cos(\phi)/h + \Pi$ where the first term describes the pressure due to a curved meniscus in the limit of classical capillaries, and $\sigma \approx$ 72 mN/m is the surface tension of water. Even for our largest channels, $P_0$ exceeds 10 bars. The second term $\Pi$ refers to the so-called disjoining pressure[1,26,27] that can reach 1 kbar at nanoscale but rapidly decreases with $h$.

For long and wide rectangular channels with $w/h$ >>1, a liquid flow driven by the pressure gradient $P/L$ is described by

$$Q = \rho(h^3/12\eta)(1 + 6\delta/h)P \times w/L \qquad (1)$$

where $\rho$ is the water density, $\eta$ its viscosity and $\delta$ the slip length. All these characteristics of nanoconfined water may depend on $h$. To find out whether eq. (1) can explain the observed behavior, we performed molecular dynamics (MD) simulations using typical parameters for water-water and water-carbon interactions (Methods). Our analysis shows that $\delta$ is large ($\approx$ 60 nm) but does not vary much with $h$ (supporting information Fig. S7), in agreement with previous MD results for flat graphene surfaces[13,14,17]. Changes in $\rho$ are also found to be relatively minor, reaching 4% for our smallest capillaries (supporting information Fig. S8). The viscosity $\eta$ increases by a factor of 2 for small $N$ < 5, which reflects the fact that water becomes more structured under nanoconfinement[28-30]. Using these parameters in eq. (1), we find that $Q$ detected for our smallest capillaries requires $P$ of the order of 1 kbar, in agreement with the magnitude of $\Pi$ expected at this spatial scale[1,27]. Our MD simulations ('Capillary pressure' in supporting information) show that, for large $N$ > 10, $P$ roughly follows the classical $P_0$ dependence with $\phi \approx$ 80° but the disjoining pressure becomes dominant at smaller $N$ reaching above 1,000 bar (supporting information Fig. S8b). Combining the simulated $P$ with the other characteristics found in our MD analysis, eq. (1) yields the $Q(N)$ dependence shown in the inset of Fig. 3b. It qualitatively reproduces our experimental findings, including the peak at small $N$ and even its absolute value. The agreement should perhaps be considered as striking, if we take into account the unresolved experiment-theory dispute[13,14,17] concerning water permeation through tubular graphene channels (CNTs).

The physics behind the non-monotonic dependence $Q(N)$ found in our MD simulations can be understood as follows. At large $N$, the classical contribution $P_0 \propto 1/h$ dominates and eq. (1) yields the linear dependence $Q \propto h$, in agreement with the trend observed in Fig. 3b for $h$ > 3 nm. Evaluating eq. (1) numerically in the classical limit ($P = P_0$), we find $Q \approx 10^{-10}$ g s$^{-1}$ µm for $h$ = 10 nm, in agreement with the values in Fig. 3b. The marked increase in $Q$ for small $N$ is due to the rapidly rising disjoining pressure whereas the final fall in $Q$ for smallest $N$ occurs due to a combined effect of decreasing $h$ and increasing $\eta$, which both reduce $Q$ overtaking the rise in $\Pi$ at small $h$. Note that, if not for the large enhancement factor $6\delta/h$ due to low friction of water against atomically smooth walls, the simulated flow would be well below our detection limit. Finally, the observed closure of monolayer capillaries ($N = 1$) seems to be not an accidental effect. Our MD analysis reveals that such narrow cavities are intrinsically unstable and collapse due to vdW attraction between opposite graphite walls (supporting information Fig. S9).

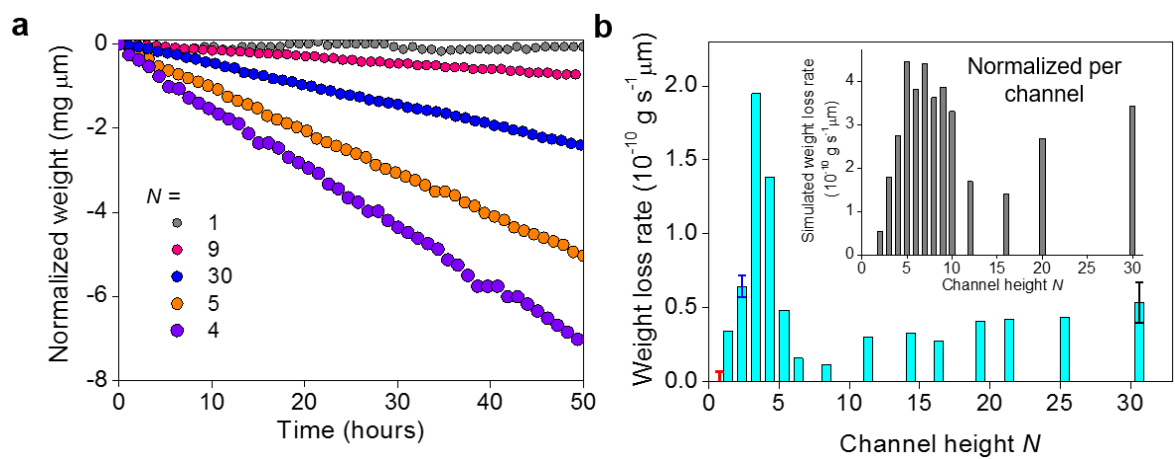

*Figure 3| Water flow through channels of different height. a, Examples of gravimetric measurements for various N. They were carried out at 21°C in near zero humidity, and the curves are normalized for devices' effective length, $\tilde{L}$. b, Dependence of Q on capillary height (the data are normalized by $\tilde{L}$ and also given per one channel). The blue error bar shows the s.d. for the data in Fig. 2b; the black bar indicates the data scatter for two devices measured for this particular height. This ±20% reproducibility is also typical for other N in which case at least two devices were measured. Only the data for N = 12 and 20 are based on single-device measurements. The red error bar indicates our detection limit (no flow could be detected for N = 1). Inset: Flow rates found in our MD simulations.*

Although the observed weight losses are determined by viscous water flow through graphene capillaries and our MD simulations describe this process well, the entire evaporation process is likely to be more complicated. In particular, the water meniscus cannot permanently stay inside channels to give rise to the capillary pressure, as assumed above. This is obvious from the fact that the observed $Q$ require a water surface area of ~1 $\mu m^2$ (Hertz–Knudsen equation), which is one-two orders of magnitude larger than the total cross-sectional area of capillaries in our devices. Accordingly, evaporation of the transported water must take place outside capillary mouths. This probably involves an atomically thin layer of water that spreads outside capillaries being driven by high spreading pressures[1]. If capillaries become full, the capillary pressure drops and the wetting layer rapidly dries up. Then the meniscus retracts back inside graphene channels leading to the next pumping cycle, in which the capillary pressure again supplies water to the surface. In this scenario, the observed weight losses are limited by a liquid flow through graphene channels, as in the experiment, whereas the actual evaporation plays only a supplementary role. Other scenarios are possible, too. For example, one can imagine that it is the spreading pressure outside capillary mouths, which drives the water flow, and its absolute value is controlled by $N$ that affects wetting film's thickness and, therefore, the driving pressure. Further work is required to fully understand the involved supplementary mechanisms.

To conclude, the demonstrated fabrication approach allows capillary devices in which the channel height can be controlled with true atomic precision by choosing spacers of different two-dimensional crystals such as graphene, boron nitride, molybdenum disulfide, etc. and their combinations. One can also alter chemical and physical characteristics of these capillaries (for example, change their hydrophilicity) using different atomically flat crystals for channel walls, which offers a large parameter space to explore. Furthermore, the availability of highly insulating materials such as boron nitride and mica allows design of nanofluidic systems in which ionic or mass transport can be controlled by gate voltage. Our current devices transfer minute amounts of liquid, typical for nanofluidics, but it is feasible to increase the flow by many orders of magnitude using dense arrays of short (submicron) capillaries covering mm-size areas, which can be of interest for, e.g., nanofiltration.

**Acknowledgments:** This work was supported by the European Research Council, the Royal Society and Lloyd's Register Foundation. B. R thanks FP7 Marie Curie fellowship.

**SUPPORTING INFORMATION**

**Making nanocapillary devices.** Our fabrication procedures are explained in Fig. S1. First, we prepare a free-standing Si nitride membrane of approximately 100×100 μm² in size using commercially available Si wafers with 500 nm thick Si nitride[31]. A rectangular hole (3×20 μm²) is made in the membrane using the standard photolithography and reactive ion etching (step 1). Then a relatively thick (> 10 nm) graphite crystal is deposited to seal the opening (step 2) using the dry transfer method described in supplementary information of ref. 32. On a separate Si wafer (with 300 nm of $SiO_2$) we prepare multilayer graphene of a chosen thickness using micromechanical cleavage[6] to serve as a spacer. The graphene crystal is patterned by electron beam lithography and oxygen plasma etching to create an array of parallel stripes of ≈ 130 nm in width and separated by the same distance (Fig. 1a). These dimensions are chosen to obtain sufficiently narrow channels (to prevent them from collapsing; see below) and, at the same time, to ensure full reproducibility using our lithography facilities. The graphene stripes are then transferred onto the bottom graphite so that they are aligned perpendicular to the long side of the rectangular opening (step 3). Oxygen plasma etching is employed to drill through the graphite-graphene stack using the hole in Si nitride as a mask (step 4). Finally, another graphite crystal (approximately, 100 nm in thickness) is 'dry-transferred' to serve as the capping layer. This completes a set of graphene capillaries, such that their entries and exits are accessible from the opposite sides of the Si wafer (step 5). After each transfer, the assembly is annealed at 400°C for 3 hours to remove possible contamination.

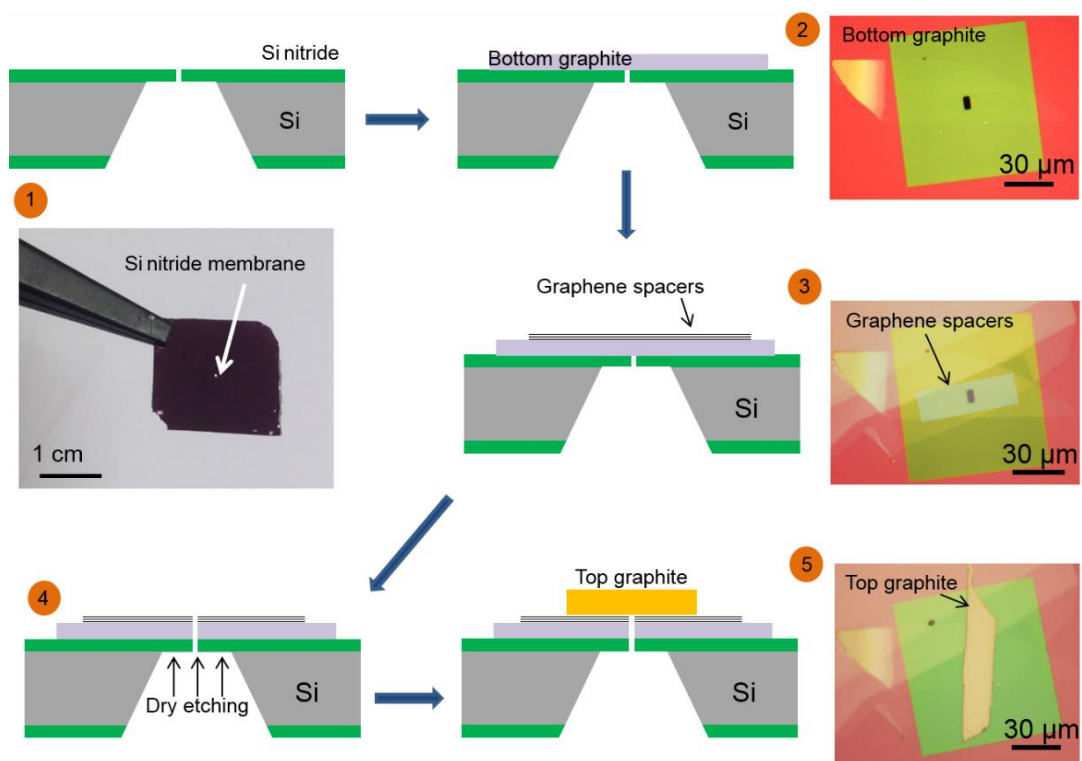

*Figure S1 | Microfabrication process flow.* *(1) A micron-scale hole is prepared in a Si nitride membrane. (2) Bottom graphite is transferred to cover the opening. (3) An array of graphene spacers is transferred on top. (4) The hole is extended into the graphite-graphene stack by dry etching. (5) Top graphite crystal is transferred to cover the resulting aperture. The accompanying optical images (in natural color) illustrate the results after each step for one of our devices. Graphene spacers are invisible in the photos and indicated by an opaque rectangle in (3). Steps 3 and 4 were often interchanged.*

**Visualization and characterization of graphene capillaries.** In addition to Fig. 1, Fig. S2 provides further examples of imaging of our graphene capillaries including their optical, atomic force microscopy (AFM), SEM and STEM micrographs. We used SEM and optical images such as in Fig. 1b and Fig. S2a to calculate the average length <L> of our devices as well as their effective average length with respect to a frictional flow, $\tilde{L}$ = <1/L>$^{-1}$. For most of our devices, their L varied by less than 30% (e.g., Fig. S2a) and, accordingly, we found no qualitative difference if using $\tilde{L}$ or <L> in our analyses.

To obtain the cross-sectional SEM images shown in the figure, we used a dual-beam system (Zeiss Crossbeam 540), which combines electron microscopy with focused ion beam (FIB) capabilities. The region of interest was located using SEM, and a protective Pt layer (≈0.5 µm thick) was deposited on top. Then a trench was milled using 30 kV Ga$^+$ beam at 0.1 nA current, which exposed the device's cross-section. Two additional polishing steps at 10 and 1 pA were subsequently carried out using the same 30 kV Ga$^+$ beam. During the final step, the raster had a width of 200 nm, and it took approximately 30 min to complete.

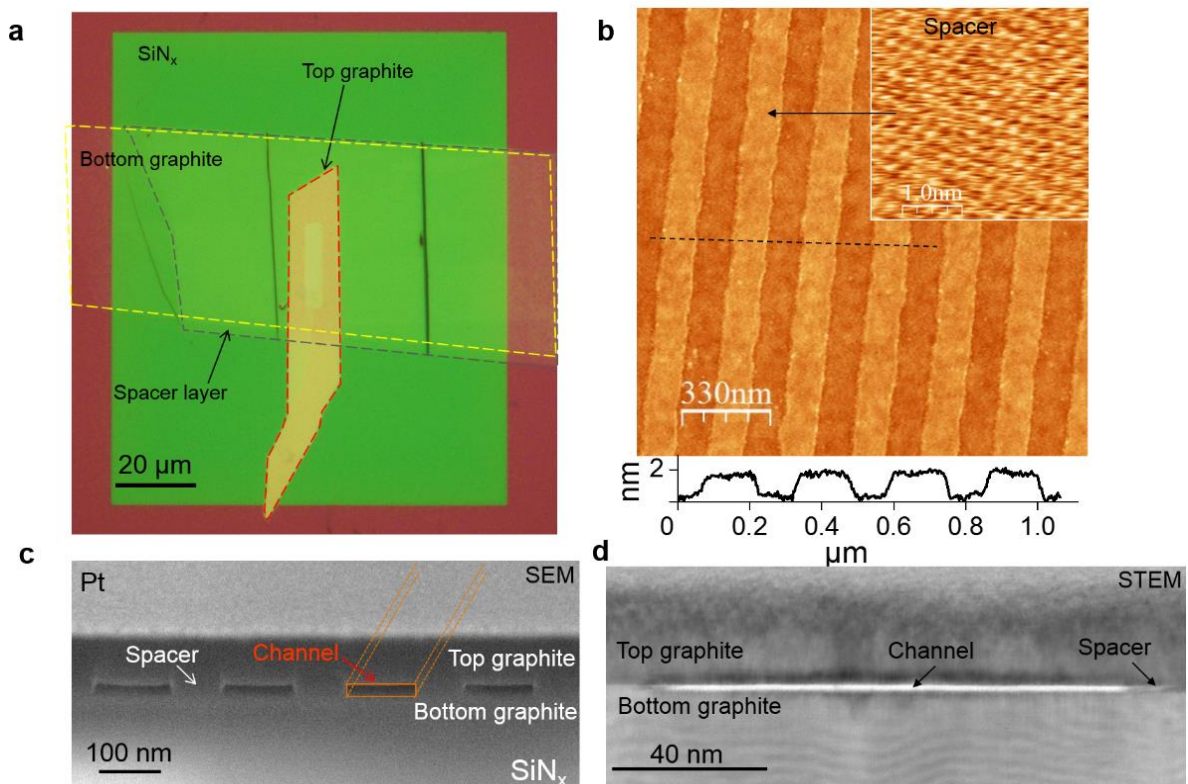

*Figure S2 | Additional images of graphene capillaries. **a,** High-magnification optical image of a final device. The green region is the free-standing Si nitride membrane. The Si wafer is seen in brown and the top graphite crystal in yellow. Red, yellow and grey contours indicate positions of the top and bottom graphite and graphene spacers, respectively. The nearly-vertical dark lines are wrinkles in the bottom layer. **b,** AFM image of 4-layer graphene spacers on top of a bottom graphite crystal (height profile along the dashed line is shown below the image). Inset: High-resolution scan (friction mode) from the region indicated by the arrow. The observation of the atomic lattice confirms that our assemblies have atomically smooth surfaces. Such smoothness is impossible to achieve using conventional materials and processes that invariably lead to the surface roughness exceeding the scale given by few-layer graphene spacers. Although the side walls of our channels are rough due to limitations of electron-beam lithography, we estimate that, because of the large ratios w/h, the side wall contribution to flow resistance cannot exceed 5% even for our 10 nm devices[34]. **c,** SEM micrograph of a capillary device with h ≈ 15 nm. **d,** Bright field STEM image of a graphene capillary with N = 4.*

Samples for STEM were obtained by implementing the *in situ* lift-out procedures[24,32,33] in a FIB system (Helios Nanolab DualBeam 660), which incorporates SEM and FIB columns to provide high-precision site-specific milling. A cross-sectional lamella (that is, a thin foil cut out perpendicular to the capillary axes) was prepared by FIB milling and lifted from the substrate using a micromanipulator, aided by ion beam deposition of platinum. After transfer to a specialist OmniProbe grid, the foil was thinned down to < 100 nm and then polished to electron transparency using 5 kV and subsequently 2 kV ion milling. High-resolution STEM images were acquired in an aberration-corrected microscope (FEI Titan G2 80-200 kV) using a probe convergence angle of 21 mrad, a HAADF inner angle of 48 mrad and a probe current of ≈80 pA. To ensure that the electron probe was parallel to the graphite planes, the cross-sectional sample was aligned to the relevant Kikuchi bands of the Si substrate and graphite.

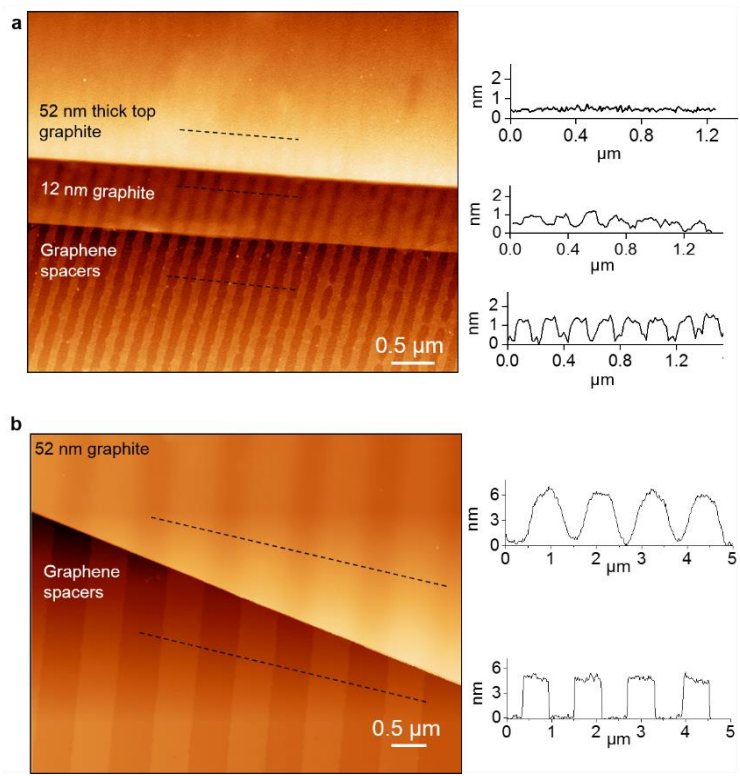

***Figure S3| Sagging of top graphite. a,*** *AFM image of trilayer channels, which are covered by a graphite layer of varying thickness.* ***b,*** *Partial sagging of the top graphite into wide channels. One can see that the top graphite bends down into the channels over their entire height h ≈ 5 nm. The insets to the right are height profiles that correspond to the traces shown by the dashed lines in the main images.*

To prevent closure of our nanocapillaries through sagging of their walls, it is essential to choose appropriate values for the channel width, $w$, and top graphite's thickness, $H$ (bottom graphite is supported by the substrate which stops it from sagging). To illustrate the crucial role of $H$, Fig. S3 shows AFM images of trilayer channels covered with graphite of a varying thickness. One can see that thin graphite ($H \approx 12$ nm) sags – at least partially – into the channels whereas the thicker layer (52 nm) remains atomically flat, which suggests that the channels underneath are likely to remain open. For our standard channels with $w \approx 130$ nm, we find that it requires $H >$ 50 nm to avoid their collapse. On the other hand, for $w \geq 500$ nm, the top graphite crystal in our experiments always sagged into the channels (even for $H > 200$ nm; Fig. S3b). It is instructive to mention that capillaries with small sagging (≤0.5 nm) reacted to high relative humidity (RH) in such a way that the sagging disappeared and the top graphite layer became flat on the AFM images. For example, for capillaries with $N = 5$ this straightening of graphene walls happened at ≈70% RH, indicating the onset of capillary condensation[1]. This allows an estimate for the contact angle $\phi \approx 55°$, in agreement with $\phi$ observed for water on clean graphite surfaces[25]. No changes with increasing RH were observed for sufficiently thick top layers that exhibited no initial sagging.

**Helium-leak testing.** To ensure that the fabricated capillaries are not blocked by sagging or contamination, we checked gas permeation through them using a helium-leak detector (INFICON UL200). A principal schematic of our experimental setup is shown in Fig. S4a. In short, a Si wafer with a capillary device is clamped between O-rings and separates two oil-free vacuum chambers. One of them is equipped with pressure gauges and a pump to allow control of the applied helium pressure $P_a$ at the capillary entry. The other chamber is connected to the leak detector. We have found our graphene-Si nitride membranes sufficiently robust to withstand $P_a$ up to 2 bar. Examples of our tests are shown in Fig. S4b. Except for devices with $N$ = 0 and 1, all other nanocapillaries allowed He permeation.

Although discussion of gas transport through graphene nanocapillaries is beyond the scope of the present report, it is instructive to compare the observed He rates $Q$ with those expected in theory. For a channel with $h$ much smaller than the mean free path $l \approx 140$ nm for He atoms at the atmospheric pressure, their mass transport is described by the Knudsen formula[35]

$$Q = \alpha \cdot P_a \cdot (M_{He}/2\pi RT)^{1/2} \cdot w \cdot h \qquad (S1)$$

where $M_{He}$ is the atomic mass of He. For narrow-slit channels, the transmission coefficient $\alpha$ can be approximated[35] by $\alpha \approx 5(h/L)$. In the case of $h$ = 15 nm and $P_a$ = 100 mbar, eq. (S1) yields $Q \approx 7\times10^{-13}$ g s$^{-1}$ µm, in good agreement with our measurements shown in Fig. S4b for $N \approx 45$. On the other hand, smaller capillaries ($N \leq 5$) are found to exhibit leak rates that are nearly two orders of magnitude higher than the rates expected from eq. (S1). Moreover, their $Q$ are even greater than those found for 10 times higher channels (Fig. S4b), contrary to general expectations. A similar enhancement of He flow was previously reported for sub-2-nm CNTs and attributed to the atomic smoothness of graphene walls[12].

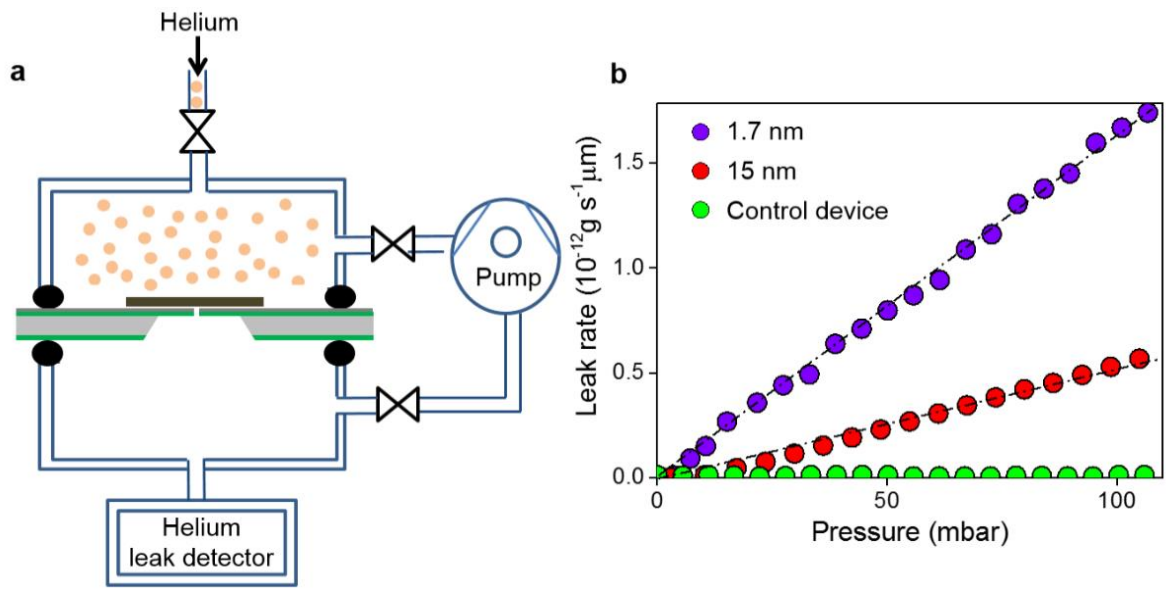

*Figure S4| He leak through graphene capillaries. a, Schematics of our setup. b, Leak rates normalized for one µm length and given per channel as a function of applied pressure for capillary devices with N = 5 and $\approx$ 45 and a control device without graphene spacers (N = 0).*

For the case of a water vapor driven by the difference in RH (23 mbar), eq. (S1) yields evaporation rates of $\approx$ 5 and 400 $\times10^{-15}$ g s$^{-1}$ µm for channels with $N$ = 5 and 45, respectively, which is 3–5 orders of magnitude smaller than the experimental values in Fig. 3b. Even if the vapor permeation is enhanced by two orders of magnitude, as observed for the He transport through capillaries with $N$ = 5, this still leaves three orders of magnitude unaccounted for. This disagreement provides yet another indication that water permeates through our graphene channels as a liquid.

**Ionic conductance.** We also tested a number of capillary devices using the electrochemical setup shown in Fig. S5a. KCl solutions of different concentrations $C$ were introduced into two reservoirs separated by a Si wafer incorporating a graphene device under investigation. Possible air bubbles were removed by extensive flushing from both sides of the Si wafer. Current-voltage (I-V) characteristics were recorded using Keithley 2636A SourceMeter and Ag/AgCl electrodes. Figure S5 shows examples of our measurements for two nanocapillary devices, with $h \approx 0.7$ and 6 nm. Devices with $N$ = 0 and 1 exhibited no detectable ionic conductance. The I-V curves are linear at low biases and exhibit little hysteresis. At high $C$, the observed ionic currents for a given voltage differ approximately by a factor of $\approx 8$, in good agreement with the ratio between the channel heights.

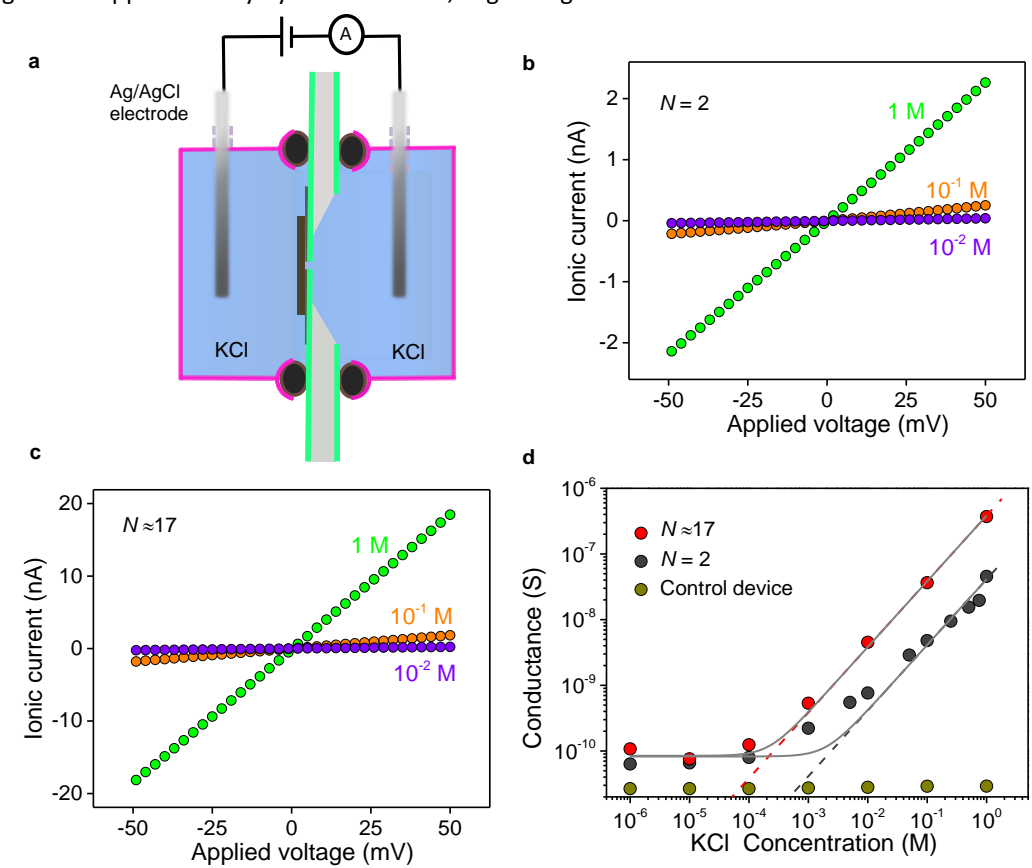

***Figure S5| Ion transport through graphene nanochannels. a,*** *Schematics of our measurement setup.* ***b,*** *Examples of I–V characteristics for the smallest capillary devices (N =2) at different KCl concentrations (L ranges from 2.8 to 7 μm).* ***c,*** *Same for a device with N $\approx$ 17 of approximately the same average length $\tilde{L}$ (L from 1.7 to 7.3 μm).* ***d,*** *Ionic conductance for these devices as a function of C (without normalizing for their slightly different $\tilde{L}$). Both blank Si nitride wafers separating the reservoirs and control devices with N = 0 (no spacers but otherwise prepared using the same fabrication procedures) exhibit leakage conductance of the order of 20 pS, which does not change with C (olive symbols). The dashed lines are G expected from the bulk conductivity of KCl for the given channel dimensions. The solid curves are fits taking into account an additional parallel conductance due to the surface charge.*

Figure S5d shows that the ionic conductance, $G$, increases linearly with $C$ for ionic concentrations higher than $10^{-2}$ M, and its absolute value agrees well with the values expected from the known bulk conductivity of KCl solutions. In the low concentration regime (< $10^{-3}$ M), $G$ saturates to a constant value, the same for both devices. Such saturation is typical for nanocapillaries and attributed to the surface charge effect[8,36]. In our case, the saturation value is very small and, taking into account electro-osmotic and finite-$\delta$ contributions[37], we find a surface charge density of $\approx 3 \times 10^{10}$ cm$^{-2}$, orders of magnitude lower than the values observed for conventional capillaries including CNTs[38]. This serves as another indication that graphene walls of our channels are impurity-free, in agreement with low charge densities usually found in graphene-based vdW heterostuctures[5].

**Gravimetric measurements.** The setup used in our studies of water transport through graphene capillaries is shown in Figs. S6a-b. The assembled capillary device was mounted on top of a container partially filled with deionized water. The container was then placed on a microbalance (Mettler Toledo XPE26) and weighted in an enclosure with a constant temperature (typically, 21±0.1°C) and at near 0% humidity that was maintained using molecular sieves. The weight of the container was recorded at regular intervals (typically, 1 min) using a computer.

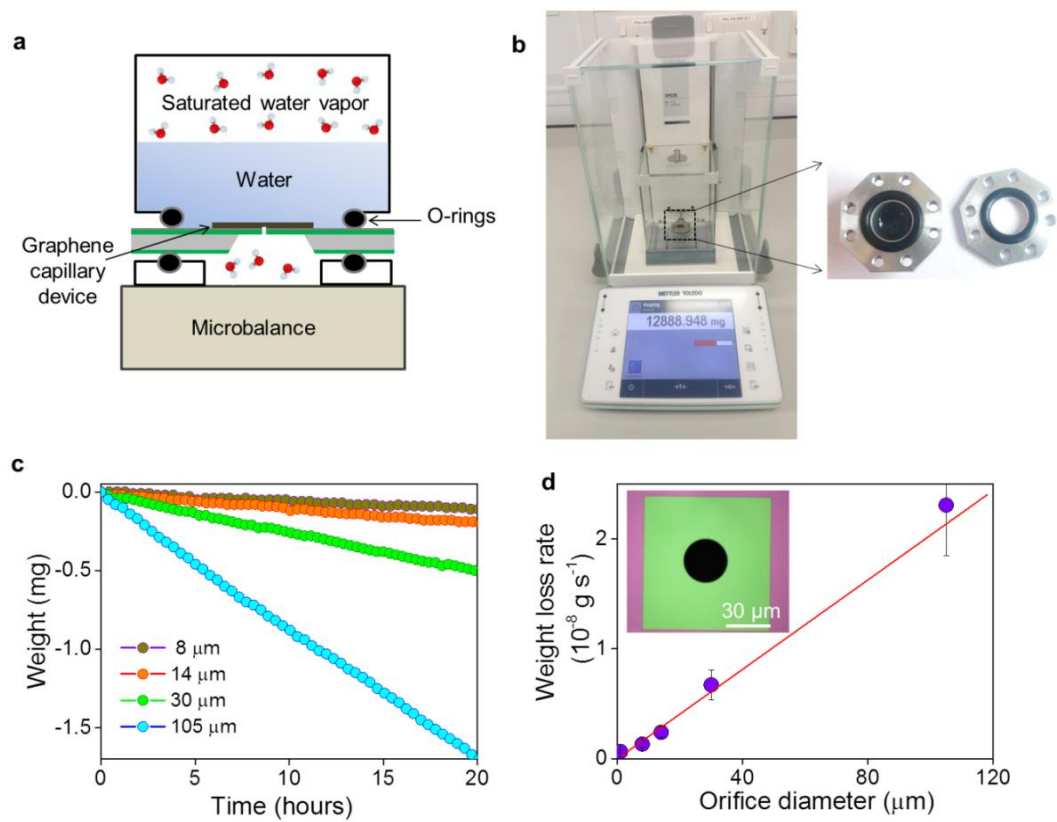

*Figure S6| Gravimetric measurements and reference devices*. *a,* Extended schematic of the experimental setup. A small aluminum container filled with water is sealed with a Si nitride wafer containing a capillary device (total weight should not exceed ~15 g to allow the required measurement accuracy). The container was weighed either upside down (water in contact with capillaries as shown in the sketch) or in the upright position as shown in the inset of Fig. 2a (capillaries are exposed to 100% RH). Both orientations resulted in the same Q. *b,* Photographs of our gravimetric setup. *c,* Examples of water evaporation through apertures of different diameters, D. *d,* Dependence of the evaporation rate on D. Red line: Best linear fit. Inset: Optical micrograph of an aperture of 30 μm in diameter, which is etched in a Si nitride membrane, seen as a green square.

To verify the accurate operation of the gravimetric setup under the same conditions as those used for our nanocapillaries, we prepared reference devices with round apertures of different diameters, $D$, etched in Si nitride membranes. Using the same sample mounting and measurements procedures as for our graphene devices, we measured water evaporation through the apertures (Fig. S6). The Knudsen numbers for our apertures are small and the evaporation can be described by diffusion of water molecules through air[35]. The molecular flow $F$ is given by[35]

$$F = 1/3 <v> l \cdot dn/dx$$

where $<v>$ is the average velocity of molecules in air, $l \approx 60$ nm is the mean free path, and $dn/dx$ the concentration gradient. To leave the container, water molecules have to diffuse through air over a distance of about $D$, which allows an estimate $dn/dx \approx \Delta n/D$ where $\Delta n = \Delta P/k_B T$ is the difference in water concentrations at large distances from the aperture and $\Delta P$ the difference in their partial pressures. The diffusion problem can be solved exactly for the case of infinitely thin orifices, which is a reasonable approximation for our 500 nm thick Si nitride membranes and yields[39]

$$dn/dx = (4/\pi) \Delta n/D.$$

The resulting weight loss is given by

$$Q = F \cdot M_{H2O} \cdot (\pi D^2/4) = <v>l \cdot (M_{H2O}/3k_BT) \cdot D \cdot \Delta P$$

where $M_{H2O}$ is the molecular weight of water. This equation yields $Q \propto D$, in agreement with the observed behavior in Fig. S6d. The counterintuitive linear dependence arises because the available area for diffusion increases proportionally to $D^2$ whereas the diffusion length decreases as $1/D$. Using $\Delta P$ = 23 mbar, the above equation yields $\approx 1.7 \cdot 10^{-10}$ g s$^{-1}$ × $D$ [in µm], which is within 15% from the best fit in Fig. S6d. Importantly, the measurements for our aperture devices cover approximately the same range of $Q$ as that found for graphene capillaries (Fig. 3). The excellent agreement between the experiment and theory confirms reliability of our gravimetry setup.

To narrow down the range of possible explanations for the observed fast water flow, two additional sets of experiments were carried out. First, using devices exhibiting fast permeation ($N$ = 5), we increased RH outside the container up to 50%. No changes in $Q$ could be detected. This shows that it was not necessary to maintain RH accurately at zero and that it is not the differential vapor pressure that drives the water flow. Most importantly, the observation indicates that evaporation from an open water surface was not a limiting factor in our gravimetry experiments. Otherwise, the increase in external humidity would significantly reduce $Q$. In the second set of experiments, we applied an additional pressure of 1.3±0.3 bar to the water column inside our containers. This pressure was chosen to be close to the maximum pressure that our membranes could withstand. To create such pressures while keeping the container weight below $\approx$15 g (required for precision gravimetry), a specified amount of NaBH$_4$ was dissolved in water inside the container which resulted in a slow release of hydrogen (over several hours at room temperature). The pressure buildup inside a closed container was monitored in a separate experiment (without a graphene device) and quantitatively agreed with the pressure expected from the chemical reaction. The extra pressure did not lead to any discernable difference in $Q$. This unambiguously proves that $P$ much higher than 1 bar push water through our capillaries. Our measurement accuracy of $\approx$10% yielded a lower bound estimate for such $P$ as 15 bar.

**Molecular dynamics simulations.** To understand the observed behavior, we used both non-equilibrium and equilibrium MD simulations (NEMD and EMD, respectively). Water molecules under a pressure of 1 bar at room $T$ were confined between two rigid graphene sheets of approximately 5×5 nm$^2$ in size and separated by $h = a \cdot N$ (Fig. S7a). Unless specifically mentioned below, we used the SPC/E model for water[40], and the carbon atoms were modeled as fixed neutral particles interacting with oxygen through the Lennard-Jones (LJ) potential with the standard values[16,41,48] of the interaction parameters, $\varepsilon_{CO}$ and $\sigma_{CO}$. For consistency, in all the presented simulations we used $\varepsilon_{CO}$ = 0.0927 kcal/mol and $\sigma_{CO}$ = 3.283 Å, and LJ interactions were truncated using a cutoff at 10 Å. The temperature of water was maintained at 300 K using the Berendsen thermostat. Long-range Coulomb forces were computed using the particle-particle particle-mesh method, and all the simulations were carried out in the canonical ensemble using LAMMPS[42]. The graphene capillary shown in Fig. S7a was initially connected to two reservoirs that contained 5,000 water molecules each. A pressure of 1 bar was applied to the water reservoirs to ensure equal pressure on water molecules inside capillaries of different $h$. Then the reservoirs were removed and periodic boundary conditions were applied in all three directions.

In our NEMD analysis, the flow was generated by applying a constant unidirectional acceleration of 10$^{12}$ m/s$^2$ to all atoms in water, which corresponds to a pressure gradient of $\approx$10$^{15}$ Pa/m. Such large gradients are standard for NEMD simulations and necessary to obtain statistically significant results[43]. The steady flow state was achieved after ~1 ns, and the data were collected for >10 ns to find the streaming velocity $V$. The flux was calculated as $Q = \rho w h V$ where $\rho$ is the average density of the nanoconfined water and $w$ the capillary width perpendicular to the flow direction. For the known $Q$, the slip length $\delta$ can be found using eq. (1). Our NEMD results are presented in Fig. S7b.

We also used EMD simulations to find $\delta = \eta/\lambda$ which is given by the ratio of the shear viscosity η to the liquid-solid friction coefficient, λ. Both η and λ were calculated through the Green-Kubo formalism using the simulated local structure of confined water[44,45]. We found that η was in the range of (0.5–0.9)×$10^{-3}$ Pa s and λ was about $10^4$ kg $m^{-2}$ $s^{-1}$, in agreement with the previous simulations for the water-graphite interface[14,43]. This yielded δ ≈ 53±8 nm for N ranging from 2 to 30 (Fig. S7b). Note that CNTs are known to exhibit a strong dependence of δ on their diameter[14], which is attributed to the effect of curvature. No h dependence was found for planar graphene channels either in our simulations or previously[14,17]. For example, Falk et al[14] reported δ ≈ 80 nm for h ranging from 0.4 to 4 nm, and Kanman et al[17] found δ ≈ 60±6 nm for h ≈ 4 nm. The relatively minor discrepancies can be attributed to details of MD simulations such as different interaction parameters, different thermostats, etc.

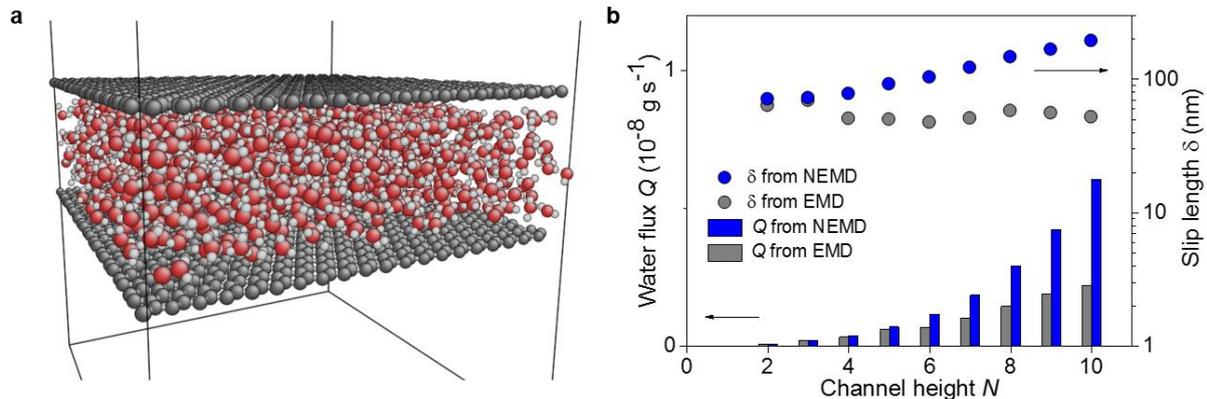

*Figure S7| Molecular dynamics simulations of water flow through graphene slits. **a**, MD setup with the simulation box indicated by the black lines. The particular snapshot is for N = 4. **b**, Simulated slip length δ and flux Q as a function of N. In the case of NEMD, δ was calculated from the simulated Q using eq. (1). Using the same pressure gradient of $10^{15}$ Pa/m for all N and δ found from EMD simulations (grey symbols), eq.(1) yields Q shown by the grey bars.*

Despite usual[43,44] quantitative differences between NEMD and EMD simulations (Fig. S7b), both show qualitatively the same behavior with a rapid decrease in water flow with decreasing N (approximately, ∝ $h^2$ as expected from eq. (1) for a constant applied pressure P) and without any anomalies at small N, in contrast to the experiment but in agreement with the previously reported simulations for flat graphene capillaries[17,43]. We also tried other models for water (TIP4P/2005)[46] and its interaction with graphene[47] as well as the use of a flexible graphene confinement. However, if a pressure P was assumed independent of h, we found it impossible to obtain a peak in permeation at small N.

**Capillary pressure.** To analyze changes in the capillary pressure P with decreasing N, we used the MD setup shown in Fig. S8a. Water molecules were supplied into graphene capillaries from a relatively large reservoir placed on the left. The reservoir was terminated with a rigid graphene sheet that was allowed to move freely from left to right. Capillary pressure sucked water inside the channel and forced the sheet to move to the right. We applied a compensating force in the opposite direction to keep the sheet stationary. From the found force and the known cross-sectional area of the channel, the pressure P was calculated.

The results are shown in Fig. S8b (solid symbols). The simulated capillary pressure rises notably faster than that expected from the classical term (red curve) due to curved menisci

$$P_0 = 2\sigma\cos(\phi)/h \qquad (S2).$$

The steeper increase in P can be understood as due to the disjoining pressure Π that consists of several contributions, including the vdW pressure $\Pi_{vdW}$ and entropic terms. The latter appear because of different densities of water inside and outside graphene nanocapillaries[16,28] as well as the enhanced structural order in nanoconfined water[29,30,41,48-50]. In our case, changes in ρ are relatively minor (inset of Fig. S8b) leading to the corresponding entropic pressure[16,28] of < 50 bar (magenta curve in the figure). Also, $\Pi_{vdW} = A/6\pi h^3$ presents a relatively small effect, where A is the Hamaker constant for water-graphite interaction[1,26]. The $\Pi_{vdW}$ contribution becomes notable only for N < 3 because of the rapid $h^{-3}$ dependence (blue curve). The total of the above three contributions is shown in Fig. S8b by the green dashed curve. The remaining difference with respect to the MD-simulated dependence can be attributed to the entropic pressure due to the increased structural order in nanoconfined water[26-30].

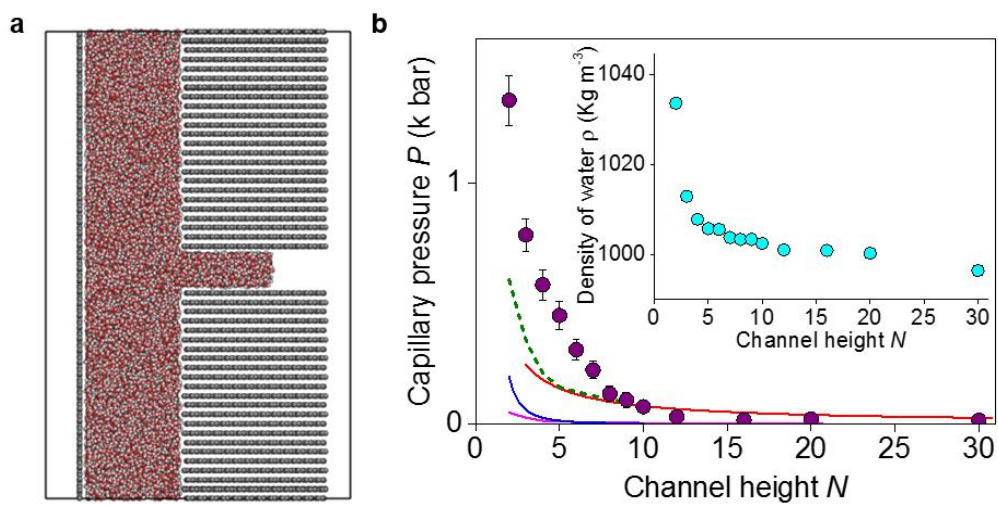

*Figure S8| MD simulations of capillary pressure. a,* Our MD setup for $N = 4$. *b,* Simulated capillary pressure (symbols with error bars). The red curve shows the best fit at large N using eq. (S2) and $\phi \approx 80°$ for water on graphite[25]. Blue and magenta curves: $\Pi_{vdW}$ with the Hamaker constant $A \approx 115$ zJ[51] and the entropic pressure due to changes in $\rho$, respectively. Green curve: Combined pressure from the three contributions. Inset: Simulated density $\rho$ of water confined between graphene sheets under external pressure of 1 bar.

**Intrinsic collapse of monolayer capillaries.** To understand the complete blockage observed for all our devices with $N = 1$, we performed the following MD simulations. Graphene capillaries were modelled as flexible graphene layers stacked on top of each other with the interlayer distance $a$. One or two graphene layers were partially removed in the middle to create channels of 20 nm in width (Fig. S9).

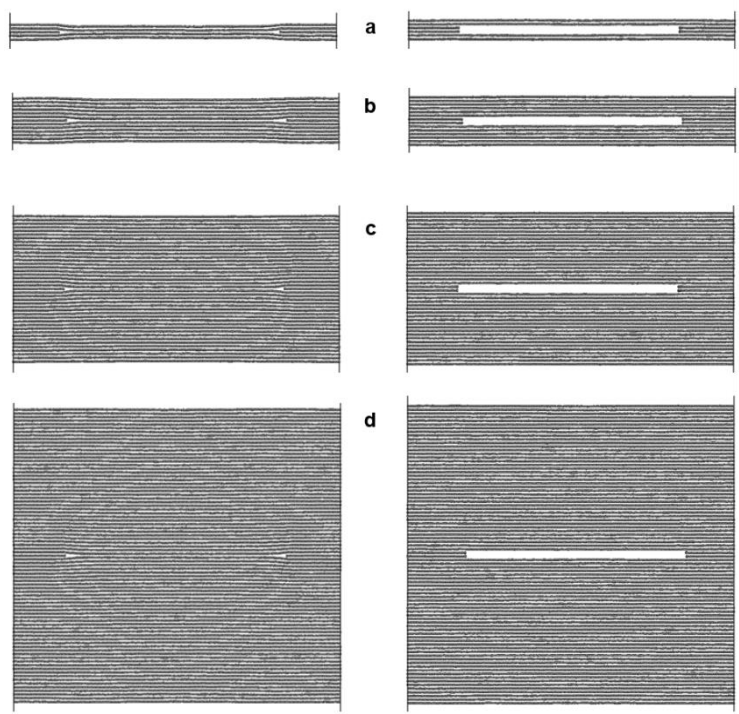

*Figure S9| Micromechanical stability of graphene cavities.* Snapshots of mono- and bi- layer capillaries (left and right columns, respectively) after 100 ps of MD simulations. *a-d,* Capillaries with different thicknesses of graphite walls. (a) to (d) are 2, 6, 20 and 40 graphene layers, respectively.

We found that the walls of monolayer channels sagged already after several ps, independent of the thickness of graphite walls. In stark contrast, bilayer channels remained open. This behavior is attributed to vdW attraction between capillary walls, which is sufficiently strong at short distances to deform the graphite bulk but rapidly vanishes with increasing the separation[52].

**Supporting information - References**